# Activation Energy of Metastable Amorphous Ge$_2$Sb$_2$Te$_5$ from Room Temperature to Melt

Sadid Muneer,[1] Jake Scoggin,[1] Faruk Dirisaglik,[1,2] Lhacene Adnane,[1] Adam Cywar,[1] Gokhan Bakan,[1,3] Kadir Cil,[1] Chung Lam,[4] Helena Silva,[1] and Ali Gokirmak[1]

[1]Department of Electrical and Computer Engineering, University of Connecticut, Storrs, Connecticut 06269, USA
[2]Department of Electrical and Electronics Engineering, Eskisehir Osmangazi University, Eskisehir 26480, Turkey
[3]Department of Electrical and Electronics Engineering, Atilim University, Ankara 06830, Turkey
[4]IBM Watson Research Center, Yorktown Heights, New York 10598, USA

Resistivity of metastable amorphous Ge$_2$Sb$_2$Te$_5$ (GST) measured at device level show an exponential decline with temperature matching with the steady-state thin-film resistivity measured at 858 K (melting temperature). This suggests that the free carrier activation mechanisms form a continuum in a large temperature scale (300 K – 858 K) and the metastable amorphous phase can be treated as a super-cooled liquid. The effective activation energy calculated using the resistivity versus temperature data follow a parabolic behavior, with a room temperature value of 333 meV, peaking to ~377 meV at ~465 K and reaching zero at ~930 K, using a reference activation energy of 111 meV ($3k_BT/2$) at melt. Amorphous GST is expected to behave as a p-type semiconductor at $T_{melt}$ ~ 858 K and transitions from the semiconducting-liquid phase to the metallic-liquid phase at ~ 930 K at equilibrium. The simultaneous Seebeck (S) and resistivity versus temperature measurements of amorphous-*fcc* mixed-phase GST thin-films show linear *S-T* trends that meet $S = 0$ at 0 K, consistent with degenerate semiconductors, and the $dS/dT$ and room temperature activation energy show a linear correlation. The single-crystal fcc is calculated to have $dS/dT = 0.153$ μV/K for an activation energy of zero and a Fermi level 0.16 eV below the valance band edge.

Chalcogenide glasses such as Ge$_2$Sb$_2$Te$_5$ (GST) have been demonstrated to rapidly and reversibly switch from the amorphous phase to the crystalline phase by annealing above the glass-transition temperature and switch from the crystalline phase back to the amorphous phase by annealing close to melting temperature followed by sudden quenching. Both the amorphous and the crystalline phases of these materials are reported to be semiconductors with a significant contrast in optical reflectivity and electrical resistivity (Figure 1a)[1], which led earlier to the use of these materials to implement rewritable optical storage media (CDs and DVDs) and more recently to nanoscale non-volatile phase change memory (PCM) devices[2–4].

PCM cells are typically constructed as two-terminal devices that are electrically switched and electrically read using pulses in the 1-100 ns time scale and can retain their data over 10 years[5]. The electrical, thermal and electro-thermal characteristics of the phase change materials used in PCM are critical for device modeling and design.

The electrical conductivity ($\sigma$) of semiconductor materials depends on the carrier mobility ($\mu$) and the free carrier concentration, $\sigma = q(\mu_n n + \mu_p p)$, where $q$ is the elementary charge, and $n, p$ are the concentrations of free electrons and holes respectively. The contributions of the mobility and the carrier concentrations can be calculated for stable unipolar materials, where conduction is dominated by either electrons or holes, using simultaneous conductivity and Hall voltage measurement[6]. However, it is not possible to melt-quench phase change materials in large scale devices the same way as the materials are amorphized in nanoscale devices. Furthermore, amorphous phase change materials are highly resistive and display a characteristic resistance drift which follows a power law behavior in time[7,8]. Hence, the measurements have to be performed at device scale with high sensitivity and be faster than the changes that take place in the amorphous material.

The free carrier mobility is typically a weak function of temperature and the carrier concentration can be a strong function of temperature, following an Arrhenius behavior $p(T) \propto \exp\left(\frac{E_A(T)}{k_B T}\right)$, where $k_B$ is the Boltzmann constant and $E_A(T)$ is the carrier activation energy, which depends on the physical processes. The activation energies reported in literature for amorphous phase change materials are calculated from slow resistivity ($\rho$) – temperature ($T$) measurements[9–11]. However, as the measurement speed is significantly slower than the rate of change in resistivity (resistance drift) of the materials, the results have to be corrected for the resistance drift as was done by Oosthoek et al.[12] Furthermore, the drifted amorphous phase change materials do not behave the same way as their (undrifted) metastable amorphous phases. Characterization of metastable and drifted amorphous as well as face centered cubic (*fcc*) phase change materials is necessary to understand the conduction mechanisms and the physical phenomena that give rise to resistance drift[10,11,13,14].

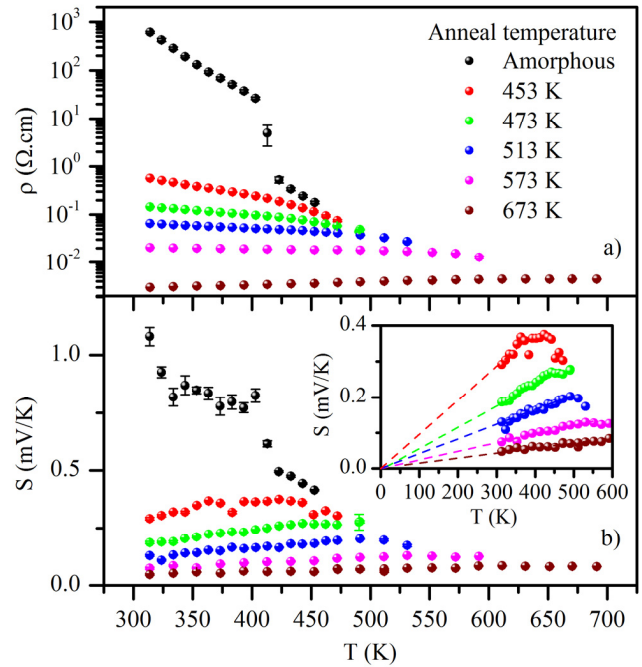

Figure 1. Simultaneous measurement of (a) resistivity, $\rho$ and (b) Seebeck coefficient, $S$ for a 200 nm thick GST film[1]. The measurement starts with amorphous GST (black markers) and cycles to increasingly maximum temperature. The legend denotes the maximum temperature to which the samples was previously annealed to. The dashed lines in the inset of (b) are linear fits made to S-T in lower temperature range meeting at $S = 0$ V/K at 0 K.

We have performed simultaneous Seebeck (*S*) and resistivity versus temperature measurements on as-deposited amorphous GST thin-films, in multiple long-duration anneal cycles (Figure 1)[1]. The films go through the expected amorphous to face centered cubic (*fcc*) and *fcc* to hexagonal closed pack (*hcp*) transitions, and finally melt and break[1,15] (Figure 2, orange line). The melting temperature ($T_{melt}$) can be considered to be the temperature at which the kinetic energy in the system is too high to maintain the crystalline structure in equilibrium[16]. Hence, the material transitions back to the amorphous phase. Disassociation of the crystals takes time, the equilibrium may be broken by substantial electronic and thermoelectric contributions[17], and mechanical stress can have a significant impact. Hence, $T_{melt}$ is not necessarily a well-defined temperature for nanoscale phase change memory devices operating at nanosecond time scales at high temperatures, high thermal-gradients, and high current densities[17–19]. We observe melting as a slight drop in the resistivity in thin film measurements if the films stay intact[1]. The thin-film resistivity measurements for $T > T_{melt}$ are not reliable as the films break apart due to surface tension.

We have reported high-speed electrical measurements on bottom contacted GST nanoscale line-cells in an earlier manuscript[8]. These measurements were performed using a waveform tailored to measure resistivity during a melting pulse with 100 ns resolution and resistivity of the amorphous phase using sinusoid segments with 48 μs resolution. The devices are cooled down to chuck temperature with a 1 μs wait time after the melting pulse when no current is forced through, to avoid filament retention and recrystallization[20], after which the resistance is again monitored using sinusoid segments with first increasing and then decreasing amplitudes. Each current and voltage segment is fitted to a sinusoid with the signal frequency and matching phase to calculate the current and the voltage values, similar to lock-in measurements. The current and voltage extracted from individual fits are used to construct the I-V hysteresis curves which show at what temperatures the material is stable during the measurement period. The resistivity of metastable amorphous GST is extracted up to 600 K, well above the glass transition ($T_{glass}$ ~ 375 K) (Figure 2, blue markers), from a large number of measurements[8].

It is not possible to directly measure the temperature of the melt during high-speed pulse measurements at device scale. Hence, $T_{device} > T_{melt}$ is considered to be achieved if resistivity is not very sensitive to temperature variations induced by increased voltage (indicated as a range in Figure 2)[8,15]. This temperature may be beyond the temperature at which the liquid semiconductor to the liquid metal transition takes place[23]; i.e. conduction and valance bands overlap. The large scale bulk melt measurement by Endo et al.[21] of liquid GST (Figure 2) possibly corresponds to the liquid metallic phase. The thin film measurements by Kato et al.[22] may correspond to the $T_{melt}$ as defined above, but be impacted by instability of the thin-films.

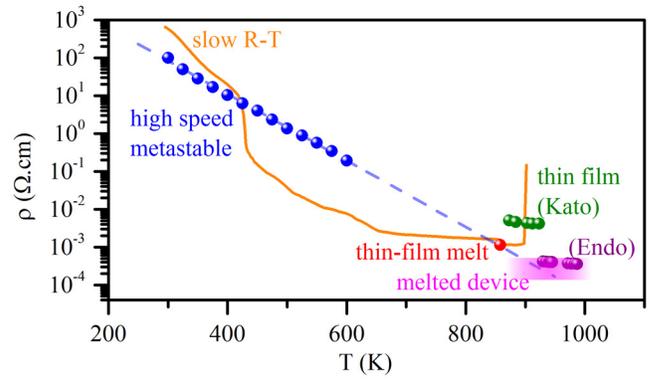

Figure 2. Device level measured metastable resistivity (blue spheres)[8] along with the thin-film resistivity (--)[1] and liquid values from the literature[21,22] as a function of T. The discrepancy between the metastable and the thin film resistivities at room temperature is likely due to the upward resistance drift of the as deposited amorphous GST. The temperature of the device-level melt resistivity is unknown (magenta box)[8,15].

The thin-film resistivity we have measured at $T_{melt}$ ($\rho_{melt}$, the red sphere in Figure 2) intersects with the extrapolation of $\rho$ versus $T$ of the device level metastable measurements which displays a simple exponential behavior (Figure 2). The $\rho_{melt}$ is lower than the resistivities in crystalline phases, even though both the *fcc* and *hcp* phases are reported to be degenerate semiconductors[24]. These observations suggest that (i) the conduction (free carrier) activation mechanisms form a continuum in a large temperature scale (300 K – 858 K), (ii) the metastable amorphous phase can be treated as a super-cooled liquid, (iii) the effective activation energies can be calculated using the linear ln($\rho$) versus $T$ behavior in this temperature range, (iv) the liquid GST continues to behave as a p-type semiconductor at $T_{melt}$ ~ 858 K, and (v) the transition from the semiconducting-liquid phase to the metallic-liquid phase of GST takes place at $T$ ~ 930 K at equilibrium[21].

Typically, the conduction activation energy ($E_A$) is calculated as the slope of the Arrhenius plot, ln($\rho$) versus $(k_BT)^{-1}$, in a narrow temperature range. Since the band-gap of amorphous GST depends on temperature and there is more than one trap level that contributes to carrier activation, calculations using ln($\rho$) versus $(k_BT)^{-1}$ yield an effective activation energy. The reported values for amorphous GST vary from $E_A$ = 0.29 eV ~ 0.33 eV[9,10] to a drift corrected $E_A$ ~ 0.39 eV in the 316 K – 343 K range[11]. The differences in the reported values may be attributed to temperature dependence of the activation energy and the impact of resistance drift, which makes the speed of the measurements and the device geometries important factors.

In this letter, we extract the effective activation energy from the metastable $\rho$-$T$ data in the 300 K – 600 K range,

where the measurements were performed faster (48 µs resolution) than the resistance drift, and our thin-film measurements at $T_{melt}$, where the amorphous phase is the stable phase. In order to calculate $E_A(T)$, we either (i) fit the $\ln(\rho)$ versus $(k_BT)^{-1}$ data with cubic splines or (ii) linearly fit $\ln(\rho)$ versus $T$ data in the whole temperature range (Figure 3). We assume that the amorphous phase has predominantly unipolar p-type conduction based on the Seebeck data[1] and the mobility is a weak function of temperature. Hence resistivity can be described using an Arrhenius behavior:

$$\rho = \rho_0 \exp\left(\frac{E_A(T)}{k_BT}\right) \quad (1)$$

where $\rho_0$ is the constant pre-factor and $E_A(T)$ is the temperature-dependent effective activation energy. Taking natural log of (1),

$$\ln(\rho) = \ln(\rho_0) + \frac{E_A}{k_BT} \quad (2)$$

and differentiating (2) with respect to $(k_BT)^{-1}$ we get,

$$\frac{d\ln(\rho)}{d(1/k_BT)} = E_A + \frac{1}{k_BT}\frac{dE_A}{d(1/k_BT)} \quad (3)$$

The cubic spline segments,

$$\ln(\rho) = a\left(\frac{1}{k_BT}\right)^3 + b\left(\frac{1}{k_BT}\right)^2 + c\left(\frac{1}{k_BT}\right) + d \quad (4)$$

where $a$, $b$, $c$, and $d$ are constants, are fit to the experimental data by matching the values and first derivatives at the connection points of the segments for the analysis. Equating the derivative of (4) with respect to $(k_BT)^{-1}$

$$\frac{d\ln(\rho)}{d(1/k_BT)} = 3a\left(\frac{1}{k_BT}\right)^2 + 2b\left(\frac{1}{k_BT}\right) + c \quad (5)$$

and (3), we get,

$$E_A + \frac{1}{k_BT}\frac{dE_A}{d(1/k_BT)} = 3a\left(\frac{1}{k_BT}\right)^2 + 2b\left(\frac{1}{k_BT}\right) + c \quad (6)$$

Hence, the closed form analytical solution of (6) is,

$$E_A = a\left(\frac{1}{k_BT}\right)^2 + b\left(\frac{1}{k_BT}\right) + c + \frac{Const}{1/k_BT} \quad (7)$$

The value of 'Const' is calculated using a known value of $E_A$ at a particular temperature. If we assume that $T_{melt}$ is the temperature at which thermal energy in the system is high enough to activate substantial number of charge carriers, the average kinetic energy of the charge carriers is approximately equal to the effective activation energy:

$$E_A(T_{melt}) = \frac{3}{2}k_BT_{melt} = 111\ meV \quad (8)$$

Using this reference, the activation energy extracted from the spline fits shows a peak value of ~377 meV at ~465 K and then a decrease for higher temperatures (Figure 3c, dashed blue line), with a room temperature value of 333 meV, comparable to previously reported values[9–11].

As an alternative approach to calculate $E_A(T)$, we have equated (1) to the experimentally observed exponential behavior (Figure 3b):

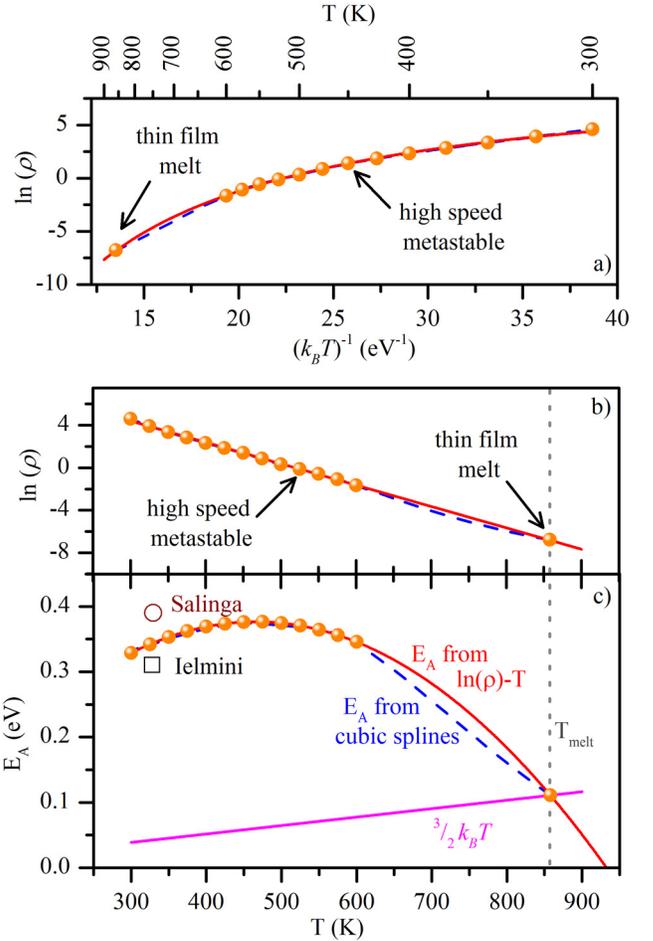

Figure 3. Natural log of measured (high speed device level) resistivity, $\rho$ (in $\Omega$.cm) as a function of 1/T (a) and T (b) (markers and lines indicate the experimental data and the fits respectively)[25]. The last data point is the thin film measurement at $T_{melt} = 858$ K[1,15]. c) Extracted activation energy by 4-point window cubic spline fits to the $\ln(\rho)$ vs. $(k_BT)^{-1}$ (--) and by linear fitting of the $\ln(\rho)$ vs. $T$ (--). The average energy of the carriers ($^3/_2\ k_BT$) (--) match $E_A$ at $T_{melt}$ (--). $E_A$ values from literature[9–11] are shown by the markers.

$$\rho = \rho_0 \exp\left(\frac{E_A(T)}{k_BT}\right) = \rho_1 e^{-\alpha T} \quad (9)$$

Hence,

$$\ln(\rho) = \ln(\rho_0) + \frac{E_A}{k_BT} = \ln(\rho_1) - \alpha T \quad (10)$$

and

$$E_A = k_BT(\ln(\rho_1) - \ln(\rho_0) - \alpha T) \quad (11)$$

where $\alpha$ and $\ln(\rho_1)$ are negative slope and intercept of the linear fit of $\ln(\rho)$-$T$ graph respectively. The only unknown in (11) is $\ln(\rho_0)$, which can be calculated from a known value of $E_A(T_{ref})$ as above:

$$\ln(\rho_0) = -\frac{E_{Aref}}{k_BT_{ref}} - \alpha T_{ref} + \ln(\rho_1) \quad (12)$$

Plugging in $\ln(\rho_0)$ from (12) to (11) yields a parabolic function for the activation energy (-- Figure 3c):

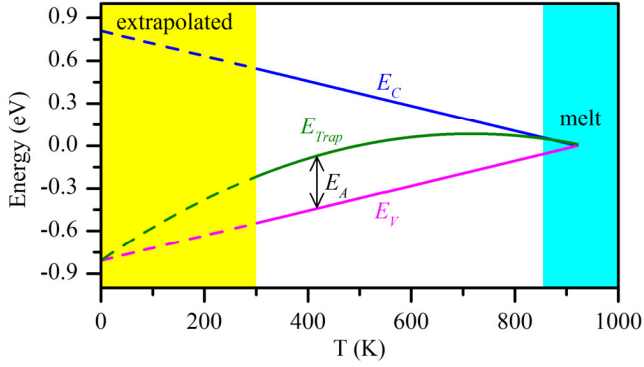

Figure 4. The conduction band edge ($E_C$), valence band edge ($E_V$), the effective trap location ($E_T$), and effective activation energy ($E_A$) constructed for metastable amorphous GST.

$$E_A = k_B T \left( \frac{E_{Aref}}{k_B T_{ref}} + \alpha T_{ref} - \alpha T \right) \quad (13)$$

The two approaches result in the same effective activation energy values in the temperature range we have experimental data. This shows that the deviations from the exponential in the experimental data, captured by the cubic splines, have a negligible effect in the calculated effective activation energies and differ only slightly in the 600 K – 855 K range. Hence, the $\rho_1 e^{-\alpha T}$ fit for the resistivity function appears to describe the resistivity behavior of the material in a wide temperature range and the parabolic $E_A(T)$ function extracted from this analysis suggests the transition to the metallic liquid phase as $E_A$ reaches zero at ~ 930 K, (Figure 3c, solid red line).

The change in activation energy for a particular trap or dopant level with temperature scales with the bandgap[26] and deeper trap levels are accessed at higher temperatures. Assuming a linearly decreasing bandgap ($E_G = E_C - E_V$) with temperature[27,28] and using the calculated $E_A(T)$, we can construct the band structure along with effective trap location, $E_{Trap} = E_A - E_V$ (Figure 4) for metastable amorphous GST. Here, $E_C$ and $E_V$ are the conduction and valence band edges respectively.

Along with amorphous GST, it is important to characterize single-crystal *fcc* GST for modeling phase change memory devices. However, it is difficult to grow large *fcc* single-crystals and GST thin-films continuously transition from amorphous to nano-crystalline *fcc* phase, staying as a mixed phase polycrystalline material until it is fully annealed above the *hcp* transition temperature. The Seebeck versus temperature behavior observed in the repeated anneal cycles shows linear *S-T* trends that meet $S = 0$ at 0 K for amorphous-*fcc* mixed phase films for various crystalline ratios[1] (Figure 1b, inset), as expected for degenerate semiconductors[22]. When we correlate $dS/dT$ to activation energy of mixed phase amorphous-*fcc* GST, $E_{A-fcc}$, for 50 nm and 200 nm thick GST films[1], we see linear trends that intersect at $E_{A-fcc} = 0$ (Figure 5). The films annealed at higher temperatures are predominantly in *hcp* phase and their resistivities increase with temperature, suggesting that there is not an appreciable change in carrier concentrations but carrier mobilities degrade as a function of temperature. The zero-activation energy means very little to no thermal energy is required to activate the carriers into the bands or over energy barriers at the grain-boundaries. Assuming that this point captures the single-crystal *fcc* GST behavior, we can model its *S-T* behavior using the constant value of $dS/dT = 0.153$ μV/K,

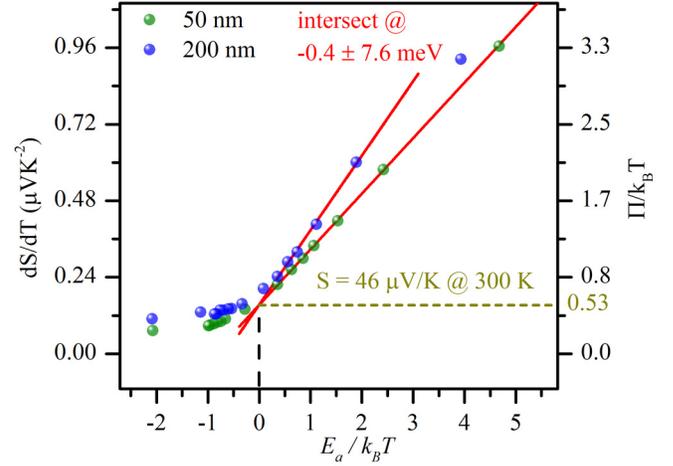

Figure 5. $dS/dT$ as a function of activation energy (at 300 K) for mixed-phase GST films annealed at different temperatures[1] and the corresponding Peltier coefficient ($\Pi$), the total energy.

$$S = \frac{dS}{dT} T = (0.153 \, \mu V/K) T \quad (14)$$

The average total energy (kinetic + potential) transported by a carrier is defined as Peltier coefficient, $\Pi = ST$, and in case of a degenerate semiconductor the energy for a hole (the majority carrier for *fcc* GST) $\Pi_h$ is approximately[18,29],

$$q\Pi_h = q \frac{dS}{dT} T^2 = q \frac{\pi^2}{3} \frac{K_B^2 T}{q} \frac{1}{E_V - E_F} T \quad (15)$$

where $E_F$ is the Fermi level. Based on this calculation, with a constant $dS/dT$ the Fermi level of single crystal *fcc* GST is 0.16 eV below the $E_V$, comparable to the previously reported location of Fermi level for *hcp* GST is ~0.1–0.2 eV below $E_V$ [22,24]. At higher temperatures, close to melting, when bipolar conduction starts, the *fcc* GST is no longer degenerate and the Fermi level is expected to deviate from this value of 0.16 eV.

In conclusion, the resistivity function of metastable amorphous GST and molten GST shows a continuum suggesting that metastable GST behaves as a super-cooled liquid, molten GST is a p-type semiconductor at $T_{melt} = 858$ K and molten GST transitions to metallic phase at ~930 K. The effective activation energy of metastable amorphous GST shows a parabolic behavior in temperature, and is expected to correspond to a distribution of trap levels that contribute to conduction. The Seebeck versus temperature data for mixed amorphous-*fcc* GST suggest that the single crystal *fcc* GST is a degenerate semiconductor having the Fermi level 0.16 eV below the valence band edge.

S. M. and J. S. were supported by DOD AFOSR MURI grant (FA9550-14-1-0351) to perform this analysis. F. D. and K. C., supported by Turkish Ministry of Education, L. A. and G. B. by DOE BES grant (DE-SC0005038), and A. C. by an NSF graduate research fellowship and NSF ECCS (#1150960) had performed the experiments.

[1] L. Adnane, F. Dirisaglik, A. Cywar, K. Cil, Y. Zhu, C. Lam, A. F. M. Anwar, A. Gokirmak, and H. Silva, J. Appl. Phys. **122**, 125104 (2017).
[2] S. W. Fong, C. M. Neumann, and H.-S. P. Wong, IEEE Trans. Electron Devices **64**, 4374 (2017).
[3] M. Wuttig and N. Yamada, Nat. Mater. **6**, 824 (2007).
[4] A. V. Kolobov, P. Fons, A. I. Frenkel, A. L. Ankudinov, J. Tominaga, and T. Uruga, Nat. Mater. **3**, 703 (2004).
[5] G. W. Burr, M. J. Brightsky, A. Sebastian, H.-Y. Cheng, J.-Y. Wu, S. Kim, N. E. Sosa, N. Papandreou, H.-L. Lung, H. Pozidis, E. Eleftheriou, and C. H. Lam, IEEE J. Emerg. Sel. Top. Circuits Syst. **6**, 146 (2016).
[6] L. Adnane, A. Gokirmak, and H. Silva, Rev. Sci. Instrum. **87**, 75117 (2016).
[7] M. Boniardi and D. Ielmini, Appl. Phys. Lett. **98**, 243506 (2011).
[8] F. Dirisaglik, G. Bakan, Z. Jurado, S. Muneer, M. Akbulut, J. Rarey, L. Sullivan, M. Wennberg, A. King, L. Zhang, R. Nowak, C. Lam, H. Silva, and A. Gokirmak, Nanoscale **7**, 16625 (2015).
[9] D. Ielmini and Y. Zhang, in *Int. Electron Devices Meet.* (IEEE, San Francisco, 2006), pp. 1–4.
[10] D. Ielmini and Y. Zhang, J. Appl. Phys. **102**, 54517 (2007).
[11] M. Wimmer, M. Kaes, C. Dellen, and M. Salinga, Front. Phys. **2**, 75 (2014).
[12] J. L. M. Oosthoek, D. Krebs, M. Salinga, D. J. Gravesteijn, G. A. M. Hurkx, and B. J. Kooi, J. Appl. Phys. **112**, 84506 (2012).
[13] M. Boniardi, A. Redaelli, A. Pirovano, I. Tortorelli, D. Ielmini, and F. Pellizzer, J. Appl. Phys. **105**, 84506 (2009).
[14] A. Calderoni, M. Ferro, and D. Ielmini, IEEE Electron Device Lett. **31**, 1023 (2010).
[15] K. Cil, F. Dirisaglik, L. Adnane, M. Wennberg, A. King, A. Faraclas, M. B. Akbulut, Y. Zhu, C. Lam, A. Gokirmak, and H. Silva, Electron Devices, IEEE Trans. **60**, 433 (2013).
[16] W. Kauzmann, Chem. Rev. **43**, 219 (1948).
[17] A. Gokirmak and H. Silva, *Crystallization and Thermoelectric Transport in Semiconductor Micro- and Nanostructures Under Extreme Conditions* (U.S. Department of Energy, 2017).
[18] G. Bakan, N. Khan, H. Silva, and A. Gokirmak, Sci. Rep. **3**, 2724 (2013).
[19] A. Faraclas, G. Bakan, L. Adnane, F. Dirisaglik, N. E. Williams, A. Gokirmak, and H. Silva, Electron Devices, IEEE Trans. **61**, 372 (2014).
[20] K. Cil, G. Bakan, Z. Jurado, Z. Woods, F. Dirisaglik, M. B. Akbulut, Y. Zhu, C. Lam, A. Gokirmak, and H. Silva, in *Mat. Res. Soc. Spring Meet.* (San Francisco, 2014), p. HH3.02.
[21] R. Endo, S. Maeda, Y. Jinnai, R. Lan, M. Kuwahara, Y. Kobayashi, and M. Susa, Jpn. J. Appl. Phys. **49**, 65802 (2010).
[22] T. Kato and K. Tanaka, Jpn. J. Appl. Phys. **44**, 7340 (2005).
[23] S. Wei, G. J. Coleman, P. Lucas, and C. A. Angell, Phys. Rev. Appl. **7**, 34035 (2017).
[24] B.-S. Lee, J. R. Abelson, S. G. Bishop, D.-H. Kang, B. Cheong, and K.-B. Kim, J. Appl. Phys. **97**, 93509 (2005).
[25] F. Dirisaglik, Ph.D. thesis, University of Connecticut, Storrs, CT, 2014.
[26] M. Kaes and M. Salinga, Sci. Rep. **6**, 31699 (2016).
[27] E. M. Vinod, R. Naik, A. P. A. Faiyas, R. Ganesan, and K. S. Sangunni, J. Non. Cryst. Solids **356**, 2172 (2010).
[28] Y. Kim, K. Jeong, M.-H. Cho, U. Hwang, H. S. Jeong, and K. Kim, Appl. Phys. Lett. **90**, 171920 (2007).
[29] N. F. Mott, *Conduction in Non-Crystalline Materials*, 2nd ed. (Oxford University Press Inc., New York, 1993) p.44.
[30] E. M. Vinod, K. Ramesh, and K. S. Sangunni, Sci. Rep. **5**, 8050 (2015).